\documentstyle[12pt]{article}
\newcommand{\nc}{\newcommand}
\nc{\beq}{\begin{equation}}
\nc{\eeq}{\end{equation}}
\nc{\beqa}{\begin{eqnarray}}
\nc{\eeqa}{\end{eqnarray}}
\nc{\lra}{\leftrightarrow}
\nc{\lmax}{l_{\rm max}}
\nc{\NCS}{N_{\rm CS}}
\nc{\betaL}{\beta_{\rm L}}

\nc{\sss}{\scriptscriptstyle}

\newcommand\lsim{\mathrel{\rlap{\lower4pt\hbox{\hskip1pt$\sim$}}
    \raise1pt\hbox{$<$}}}
\newcommand\gsim{\mathrel{\rlap{\lower4pt\hbox{\hskip1pt$\sim$}}
    \raise1pt\hbox{$>$}}}

\begin{document}

\input epsf.tex
\title{Classical Sphaleron Rate on Fine Lattices}
\author{Guy D. Moore\footnote{email:  guymoore@physics.mcgill.ca} ${}^{,a}$
	 and Kari Rummukainen\footnote{email:  kari@nordita.dk} ${}^{,b,c}$
\vspace{2cm} \\
${}^a$ {\small Dept. of Physics, McGill University, 3600 University St.}\\
        {\small Montreal, QC H3A 2T8 Canada}\\
${}^b$ {\small NORDITA, Blegdamsvej 17, DK-2100 Copenhagen {\O} , Denmark}\\
${}^c$ {\small Helsinki Institute of Physics,
               P.O.Box 9, 00014 University of Helsinki, Finland}\\
}

\maketitle
\begin{abstract}
We measure the sphaleron rate for hot, classical Yang-Mills theory on
the lattice, in order to study its dependence on lattice spacing.  By
using a topological definition of Chern-Simons number and going to
extremely fine lattices (up to $\beta=32$, or lattice spacing $a = 1 /
( 8 g^2 T )$) we demonstrate nontrivial scaling.  The topological
susceptibility, converted to physical units, falls with lattice spacing
on fine lattices in a way which is consistent with linear dependence on
$a$ (the Arnold-Son-Yaffe scaling relation) and strongly disfavors a
nonzero continuum limit.  We also explain some unusual behavior of the
rate in small volumes, reported by Ambj{\o}rn and Krasnitz.

\end{abstract}

\vskip1cm
\leftline{}
\leftline{MCGILL-99/21}
\leftline{NORDITA-99/33HE}
\leftline{June 1999}

\vskip-19cm
\rightline{}  
\rightline{MCGILL-99/21}
\rightline{NORDITA-99/33HE}
\rightline{hep-ph/9906259}

\newpage

\section{Introduction}

Baryon number is not a conserved quantity in the standard model.
Rather, because of the anomaly, its violation is related to the
electromagnetic field strength of the SU(2) weak
group \cite{tHooft},
\beq
\partial_\mu J^\mu_B = N_G \frac{g^2}{32 \pi^2} 
	\epsilon_{\mu \nu \alpha \beta} {\rm Tr} F^{\mu \nu} 
	F^{\alpha \beta} = N_G \frac{g^2}{8 \pi^2} E_i^a B_i^a \, ,
\eeq
where $N_G=3$ is the number of generations.\footnote{There is also a
contribution from the hypercharge fields, but it will not be relevant
here because the topological structure of the abelian vacuum does not
permit a permanent baryon number change.}
The right-hand side of this equation is not surprisingly a total
derivative, with an associated charge called the Chern-Simons
number, 
\beqa
\frac{N_B}{N_G} = \frac{1}{N_G} \int d^3 x J^0_B & = & \NCS \equiv 
	({\rm integer}) + \frac{g^2}{32 \pi^2} 
	\int d^3 x \epsilon_{ijk} \left(
	F_{ij}^a A_k^a - \frac{g}{3} f_{abc} A^a_i A^b_j A^c_k 
	\right) \, , \\
\NCS(t_1) - \NCS(t_2) & \equiv & \int_{t_2}^{t_1} dt \int d^3 x 
	\frac{g^2}{8 \pi^2} E^a_i B^a_i \, .
\label{def1_NCS}
\eeqa
Chern-Simons number $\NCS$ has topological meaning; its change in a 
vacuum to vacuum
process is the (integer) second Chern class of the gauge connection.
Note that, according to the first equation, the total baryon number
$N_B$ need only be an integer in vacuum, when $\NCS$ is an integer.  The
baryon number in vacuum fixes the constant of integration in the
definition of $\NCS$.

In vacuum, baryon number can be violated by a vacuum fluctuation large
enough to have a nonzero integer Chern-Simons number.  The efficiency of
baryon number violation by this mechanism is totally negligible
\cite{tHooft}; but at a sufficiently high temperature baryon number
violation can proceed by thermal excitations which 
change $\NCS$, and the rate for such a process is not necessarily very
small \cite{ArnoldMcLerran}.  This can
have very interesting cosmological significance, since it complicates
GUT baryogenesis mechanisms and opens the possibility of baryogenesis
from electroweak physics alone.  This motivates a more careful
investigation of baryon number violation in the standard model at high
temperatures.  

The baryon number violation rate relevant in cosmological settings can
be related by a fluctuation dissipation relation
\cite{KhlebShap,Mottola,RubShap} to the
diffusion constant for Chern-Simons number,
\beq
\label{def_of_Gamma}
\Gamma \equiv \lim_{V \rightarrow \infty} \lim_{t \rightarrow \infty} 
	\frac{\langle ( \NCS(t) - \NCS(0) )^2 \rangle}{Vt} \, ,
\eeq
where the angular brackets $\langle \rangle$ represent a trace over the
thermal density matrix.  This quantity is called the sphaleron rate for
historical reasons.  There has been some controversy not only to its
size in the symmetric electroweak phase (which closely resembles pure
Yang-Mills theory, an approximation we will make from here on) but even
to its parametric dependence.  On purely dimensional grounds, at high
temperature it must scale as $T^4$, but the dependence on the coupling
constant has been more controversial.  Since the natural nonperturbative
length scale of hot Yang-Mills theory at weak coupling is $\sim 1 /
( g^2 T)$ it was long believed that $\Gamma \propto \alpha_w^4 T^4$.  In
this case, $\Gamma$ could equal its value in classical Yang-Mills
theory, as originally suggested by Grigoriev and Rubakov
\cite{GrigRub}. 

More recently Arnold, Son, and Yaffe (ASY) have argued \cite{ASY}
that while the natural
length scale for hot, weakly coupled Yang-Mills theory is $\sim 1 /
( g^2 T)$, the natural time scale is different; because of interactions
between the nonperturbative infrared excitations and essentially
perturbative but very numerous UV excitations, the time evolution of IR
Yang-Mills fields should be overdamped and the natural time scale for
their evolution should be $\sim 1 / ( g^4 T)$, or, restoring $\hbar$,
$\sim 1 / (\hbar g^4 T)$.  The appearance of $\hbar$ in this expression
precludes any simple correspondence between the classical theory and the
quantum theory.  They then argue that, since only the nonperturbative IR
fields can contribute to the diffusion of $\NCS$, the correct parametric
behavior for $\Gamma$ is $\Gamma \propto \alpha_w^5 T^4$.  The analogous
expression in the classical theory has some UV regulator serving the
role of $\hbar$ in the quantum theory; for instance, for Yang-Mills
theory on a lattice under the standard lattice action, the UV regulator
scale is the inverse of the lattice spacing $a$, so the natural time
scale should be of form $1 / ( g^4 a T^2)$, leading to $\Gamma \propto a
\alpha^5 T^5$ \cite{Arnoldlatt}.  Subsequently, B\"{o}deker showed that
the coefficient of the $\alpha^5$ law should contain a further
logarithmic dependence on $g^2$ \cite{Bodeker}, or, on the lattice, on
$g^2 aT$.

The arguments of ASY are still considered somewhat
controversial.  While numerical simulations of 
classical Yang-Mills theory, supplemented
with added degrees of freedom intended to serve the role of the ``hard''
quantum UV degrees of freedom, clearly support their arguments
\cite{particles}, results for $\Gamma$ in pure classical Yang-Mills
theory on the lattice \cite{AmbKras,slavepaper,AmbKras2} 
have never convincingly displayed linear scaling in lattice spacing.
Furthermore, the more recent work of Ambj{\o}rn and Krasnitz finds two
other results which are problematic to interpret if Arnold, Son, and
Yaffe are correct; they find overly strong lattice spacing dependence
for $\Gamma$ in small volumes, and they find unexpectedly rapid falloff
for unequal time, Coulomb gauge fixed correlators \cite{AmbKras2}.

We will not address the question of unequal time, Coulomb gauge
correlators here, except to note that we believe such correlators should
show a strong {\em volume} dependence even at fixed $k$.
Since the physical volume was varying along with the
lattice spacing in Ambj{\o}rn and Krasnitz' results, it is difficult to
disentangle these two dependencies.  We leave settling this problem to
future work, but it is our general belief that such correlators will not
prove very useful probes of infrared dynamics.

This work is intended to answer the other two questions about the
applicability of the ASY picture to classical Yang-Mills theory.
First, we present  results for $\Gamma$ on a much wider range of lattice
spacings, including much smaller spacings $a$, than has previously been
done.  Although the results at fairly large $a$ show weak lattice
spacing dependence, as $a$ becomes smaller a strong $a$ dependence sets
in,  which is incompatible with $\alpha^4$ scaling, but is in 
accordance with expectations if Arnold, Son, and Yaffe are correct.  We
also present a re-analysis of the behavior of $\Gamma$ in a small, fixed 
volume.  When care is taken to ensure that the physical volume really
remains fixed as the lattice spacing is varied, we find $\Gamma$ to
depend on $a$ slightly more weakly than in large volumes.  This is also
expected in the ASY picture.

\section{Expected scaling behaviors}

To explain the different proposed scaling behaviors for $\Gamma$, 
we will briefly
review the thermodynamics of classical Yang-Mills theory.
The partition function for classical Yang-Mills theory is equivalent to
the path integral for three dimensional, quantum Yang-Mills theory with
an added adjoint scalar, as originally shown by Ambj{\o}rn and Krasnitz
\cite{AmbKras}.  The classical Yang-Mills partition function,
absorbing $g$ into the definition of the connection so covariant
derivatives are $D_i = \partial_i + i A_i$, is
\beqa
Z & = & \int {\cal D}A_i^a {\cal D} E_i^a \delta((D_i E_i)^a) 
	\exp( - H/T ) \, , \\
H & = & \int d^3 x \left( \frac{1}{4g^2} F_{ij}^a F_{ij}^a 
	+ \frac{1}{2} E_i^a E_i^a \right) \, .
\eeqa
The delta function enforces Gauss' law.  It is convenient to rewrite
it by introducing a Lagrange multiplier,
\beq
\delta((D_i E_i)^a) = \int {\cal D}A_0^a \exp \left( 
	\frac{i}{g} \int A_0^a (D_i E_i)^a \right) \, ,
\eeq
where $A_0$ has the same normalization as $A_i$.
The integral over $E$ is then Gaussian.  Performing it yields the
partition function
\beqa
Z & = & \int {\cal D}A_i {\cal D}A_0 \exp(-H/T) \, , \\
H/T & = & \int d^3x \left( \frac{1}{4 g^2T} F_{ij}^a F_{ij}^a + 
	\frac{1}{2 g^2T} (D_i A_0)^a (D_i A_0)^a \right) \, .
\label{partition_func}
\eeqa
Physically the $A_0$ field corresponds to the time component of the
connection, but if we choose to interpret the result as a path integral
for a 3-D quantum field theory then the $A_0$ field just corresponds to
some massless adjoint scalar.  

The form for the partition function coincides, except for the absence of a
mass term for the $A_0$ field, with the path integral for the full
quantum theory in the dimensional reduction approximation
\cite{KRS,FKRS,KLRS}, and so it quite accurately reproduces the
thermodynamics of infrared Yang-Mills fields.  This motivates the belief
that the dynamics of the full Yang-Mills theory will also coincide with
the dynamics of the classical theory.  Looking carefully at
Eq. (\ref{partition_func}), we observe that $g^2$ and $T$ only enter in
the combination $g^2 T$.  Now $\Gamma$ has engineering dimension 4, so
since $T$ is the only dimensionful quantity in hot Yang-Mills theory at
weak coupling, we must have $\Gamma \propto T^4$.  The ``naive'' scaling
argument for $\Gamma$ corresponds to requiring that each $T$ carry a
factor of $g^2$, as is motivated by the form of the partition function.  

The problem with this argument, and with the argument that the
dynamics of the classical and quantum theories should correspond, is 
that it neglects the effects of UV divergences.  For thermodynamic
quantities the UV divergences of the classical theory arise from a
finite number of graphs and can be absorbed by counterterms.  However,
from the way that the $A_0$ field arose above, the classical theory is
obliged to have a zero {\em bare} mass squared for this field.  Hence the
{\em physical} mass squared will approximately equal the linear one loop
divergent contribution, 
$m_D^2 \propto g^2 T \Lambda$, with $\Lambda$ some UV cutoff.  In
particular, on the lattice and using the conventional (Wilson) action,
$1/a$ serves the role of $\Lambda$, and \cite{KRS}\footnote{The result
in the reference differs by a factor of 5/4 because it includes a
contribution from a Higgs field, which is absent here since we treat
pure Yang-Mills theory.}
\beq
\label{mD_is}
m_D^2 = \frac{\Sigma g^2 T}{\pi a} \, , \qquad
	\Sigma = 3.175911536 \ldots \, .
\eeq
The thermodynamics of the
gauge fields $A_i$ do not care about this divergence, since their
thermodynamics have a finite limit as $m_D^2 \rightarrow \infty$.
Therefore the argument that the only natural length scale for the $A_i$
fields is $1 / g^2 T$, remains valid in the presence of UV divergences.
However, since the dynamics of the $A_i$ and $A_0$ fields are
intertwined, it is not at all clear that unequal time phenomena
involving the $A_i$ fields should be UV regulation insensitive.

Arnold, Son, and Yaffe \cite{ASY} examined the propagator of the full
quantum theory at soft momentum, including the UV influences at leading
order in the coupling by including the hard thermal loop (HTL)
self-energy contribution \cite{HTL}.  For spatial momenta $p$ in the
parametric regime $p \ll gT$ they conclude that the evolution of the
transverse (magnetic) degrees of freedom responsible for baryon number
violation is overdamped, of form
\beq
\frac{dA(p)}{dt} = - \frac{4 p^3}{\pi m_D^2} A(p) + ({\rm noise}) \, .
\label{overdamp}
\eeq
The perturbative treatment which gives this result holds provided that
$p \gg g^2 T$, but it breaks down when $p$ is of order $g^2 T$, the
scale which we are interested in.  Ignoring this difficulty and applying
it for $p \sim g^2 T$, we
find that the time scale for significant change in $A$ is $\sim m_D^2 / g^6
T^3$, which in the continuum theory is $\sim 1 / (g^4 T)$ and on the
lattice is $\sim 1 / (g^4 a T^2)$.

Of course it is problematic to apply Eq. (\ref{overdamp}) beyond its
range of applicability.  However, for $g$ sufficiently small, we may
apply it at the scale $p \sim g^{2-\epsilon} T$ to find that the natural
time scale here is $(1/t) \sim g^{4-3 \epsilon}T$.  Since the time scale
for field evolution at $p \sim g^2 T$ cannot be faster than that at a
larger value of $p$, Eq. (\ref{overdamp}) does place a bound on the
natural time scale for fields with $p \sim g^2 T$; the time scale 
cannot differ from
the estimate $\sim 1 / (g^4 T)$ by a nonzero power of $g$.  Thus,
Eq. (\ref{overdamp}) is enough to ensure the ASY result up to
corrections weaker than any power of $g$.  A much more careful
study of the HTL effective theory by B\"{o}deker \cite{Bodeker,Bodeker2}
finds that $\log(1/g)$ corrections do occur; however, they prove to be
numerically small \cite{Bodek_paper}.

Note however that the ASY argument is a parametric treatment which
relies on a large separation between the $gT$ and $g^2 T$ scales.
Numerical results, for instance for the subleading contributions to the
Debye screening length \cite{LainePhilipsen}, suggest that $g^2$ (or, on
the lattice, $g^2 a T$), may need to be fairly small before such a
separation of scales really exists.  Therefore, we might expect that the
ASY scaling behavior only sets in at reasonably small lattice spacings
$a$.  This motivates the study of the sphaleron rate on very fine
lattices, which we take up in the next section.

\section{Results:  large lattices, fine spacings}

To address the question: ``How does $\Gamma$ depend on lattice spacing
$a$ when $a$ is small?'' we track $\NCS$ on the lattice.
We use the conventional Kogut-Susskind lattice action
\cite{Kogut}, which is the Minkowski time analog of the Wilson
action for Euclidean lattice Yang-Mills theory.  We use the same 
discrete implementation for the time 
evolution as in Ambj{\o}rn et. al. \cite{Ambjornetal} and almost all
subsequent work.  Besides the size of the lattice there is one variable,
$\betaL$, which is a reciprocal temperature in lattice units.  At tree
level it is $\betaL = 4 / ( g^2 a T)$, but this relation receives
radiative corrections because the UV lattice and continuum fields behave
differently.  These are treated, for the case $m_D^2 \ll 1/a^2$, in
\cite{Oapaper} and extended to larger $m_D$ in Appendix B of
\cite{particles}.  We use
the expression from there for the one loop improved relation between
$\betaL$ and lattice spacing $a$, 
\beqa
\label{latt_improvement}
\betaL & = & \frac{4}{g^2 aT} + \left( \frac{1}{3} 
	+ \frac{37 \xi}{6 \pi}
	\right) - \left[ \left(\frac{4}{3} + \frac{2 a^2 m_D^2}{3}
	+ \frac{a^4 m_D^4}{18} \right) \frac{\xi(am_D)}{4 \pi} 
	- \left( \frac{1}{3} + \frac{a^2 m_D^2}{18} \right) 
	\frac{\Sigma(a m_D)}{4 \pi}
	\right] \nonumber \\
\betaL & \simeq & \frac{4}{g^2 aT} + 0.6 \qquad {\rm or}
	\qquad a \simeq \frac{4}{(\betaL - 0.6) g^2 T}\, ,
\eeqa
where $\xi=0.152859 \ldots$ and the functions $\xi(x)$ and $\Sigma(x)$
are defined in \cite{particles}; the approximation that the correction
term is $0.6$ is good to $10\%$ for all $a$ considered here but we will
use the full expression.  In this section the difference between
the naive and improved match will not be important, but in the next
section it is essential.  Henceforth, when we refer to $\betaL$ we will
mean the ``unimproved'' quantity appearing in
Eq. (\ref{latt_improvement}); while the quantity $\beta$ will mean
$\beta \equiv 4 / (g^2 aT)$ using the improved relation for $a$.

The measurement of Chern-Simons number deserves some comment.  
Early work on Chern-Simons number diffusion 
\cite{Ambjornetal,AmbKras,Moore1,TangSmit} used ``naive''
definitions of $\NCS$ in which the right hand side of
Eq. (\ref{def1_NCS}) is implemented as the integral of a local operator
over the unmodified lattice fields.  This approach is not topological
because such a local operator on the lattice is never a total
derivative, as it should be for $\NCS$ to have topological meaning.
Because of this, such a definition of $\NCS$ shows diffusive behavior
even when there is no true diffusion of baryon number occurring
\cite{MooreTurok,ASY}.  The response to true topology change can also
get renormalized by UV fluctuations.  The latter problem gets less
severe as $a$ is reduced, but we find that the
``spurious'' diffusion per physical 4-volume grows as $a^{-1}$ and is
therefore disastrous.  Therefore we should use a topological definition
of $\NCS$.

Technically topology is not well defined for lattice
fields, but it can be well defined on a restricted class of lattice
fields which are sufficiently ``smooth''
\cite{Luscher,Woit,Phillips}.  In our context topology is well
defined for suitably small lattice spacing; in practice there is no
problem if $\beta \geq 6$ $(a < 2 / 3 g^2 T)$, which will be the case
for almost all lattices we consider.  For the current application, two
topological means have been developed; the ``slave field'' method
\cite{slavepaper}, similar to the method of Woit \cite{Woit}; and
``calibrated cooling'' \cite{broken_nonpert}, an improvement on the
field smearing proposal of Ambj{\o}rn and Krasnitz \cite{AmbKras2}.  The
``slave field'' method is numerically efficient but noisy, which means
in practice that a longer numerical evolution is needed to get good
statistics.  Therefore we will use ``calibrated cooling''.  The
philosophy of the method is that the topological content of the
connection cannot be modified by small local changes; therefore we may
``smear'' the connection, removing UV noise which is responsible
for the poor performance of the ``naive'' measurement method.  After the
smearing we measure $\NCS$ by integrating an $O(a^2)$ improved local
operator for $E_i^a B_i^a$, and we cure the slight residual error in the
algorithm by ``calibrating'' it with occasional coolings all the way to
vacuum; the full details can be found in \cite{broken_nonpert}.

\begin{figure}[t]
\centerline{\epsfxsize=3in\epsfbox{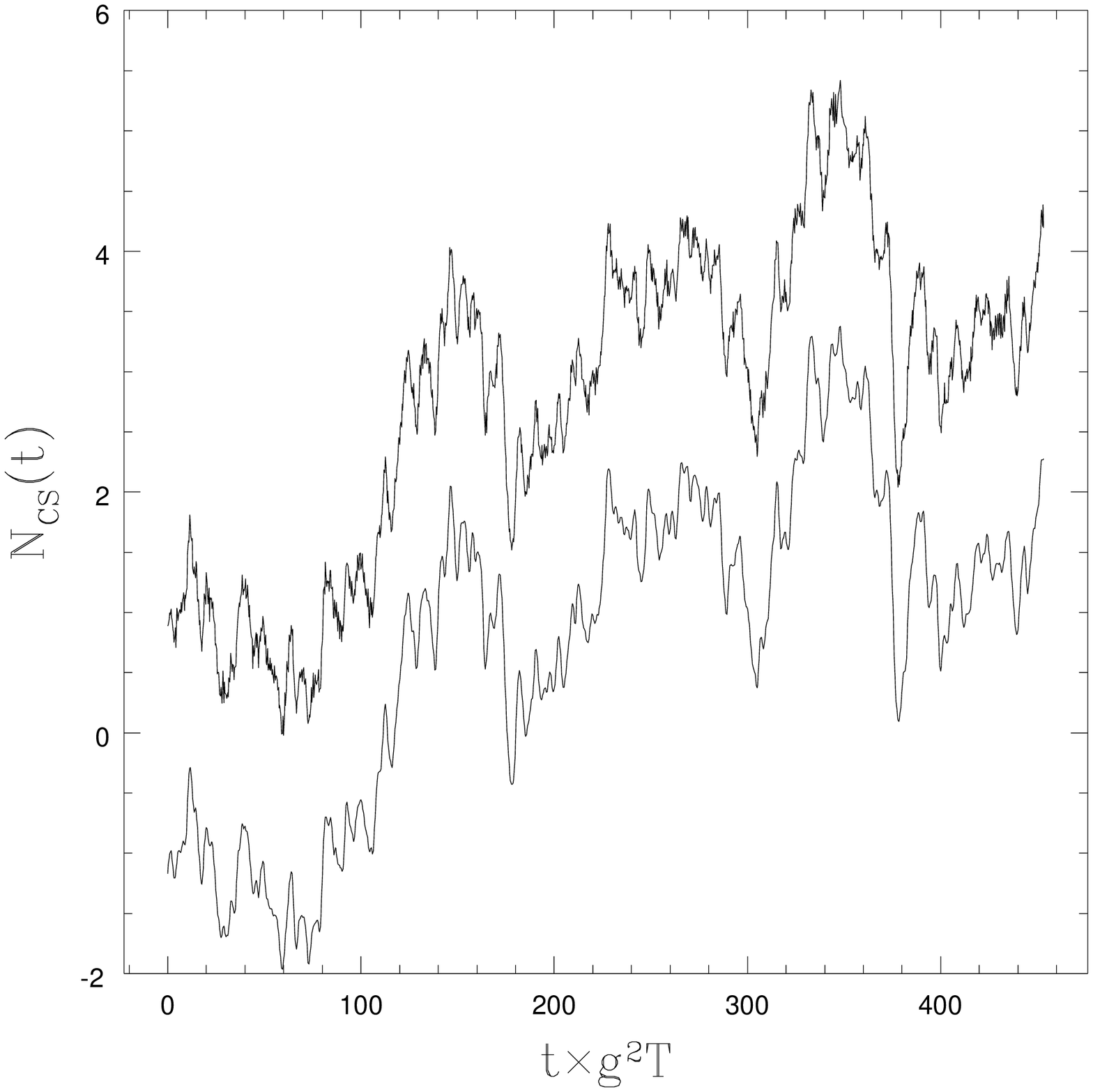} \hspace{0.3in} 
\epsfxsize=3in\epsfbox{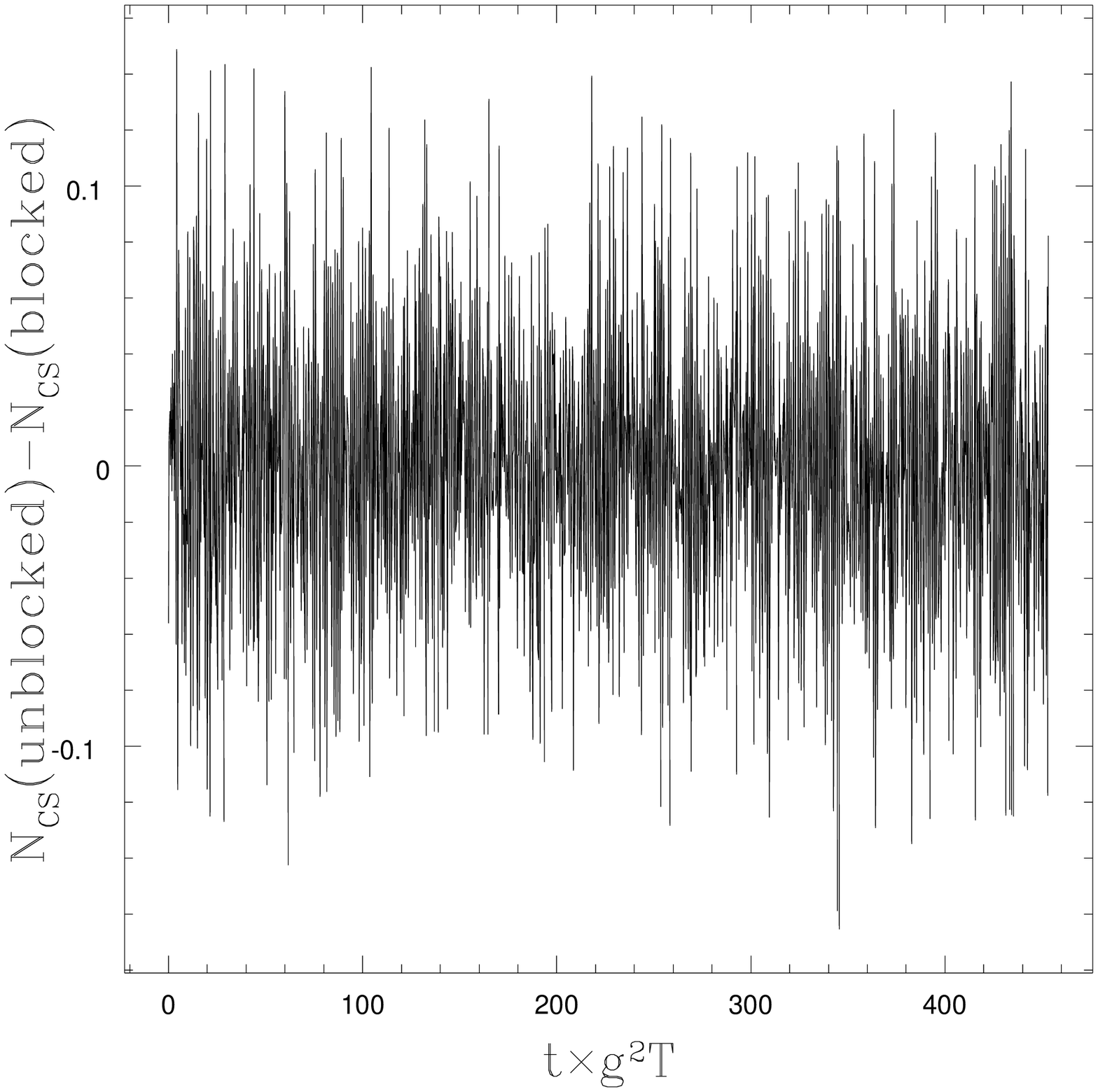}}
\caption{\label{testblock} Right:  $\NCS$ measured with and without 2-fold
blocking on a $32^3$ lattice at $\beta=12$ ($a=1/3 g^2T$).  The smearing 
depth is the same, $7.5 a^2=0.83/g^4 T^2$ for the two curves, with
blocking (upper curve) and without (lower curve, shifted down for
clarity).  Left:  the difference between the curves, which 
is small and spectrally
white; note the difference in scale between the plots.}
\end{figure}

The problem with the above approach is its numerical cost; measuring
$\NCS$ as a function of time is much more expensive than generating the
Hamiltonian trajectory.  This is only really a problem on very fine lattices,
as numerical cost rises as $a^{-4}$.  However, in this case the topology
changing configurations we are after are many, many lattice spacings
across and are highly over-resolved; no topological information is lost
by ``blocking'' the lattice and then measuring $\NCS$.  That is, we
construct a ``blocked'' lattice with $B$ 
times the original lattice spacing by making the
connection between neighboring points on the blocked lattice equal to
the product of links between the same points on the unblocked lattice;
the details are in \cite{broken_nonpert}.  We do this first, before
applying any smearing.  This introduces some white noise into $\NCS$ but
does not change the diffusion at all, which means that results for
$\Gamma$ will be unchanged.  Figure \ref{testblock} shows a test of
blocking in which we track $\NCS$ for the same Hamiltonian trajectory,
with and without blocking.  The difference is small and purely
spectrally white.
On the other hand, the difference in numerical effort is enormous;
without blocking, measuring $\NCS$ took 5 times as much CPU time as
updating the fields, while with blocking it took around $10\%$ as much.
We will block for all data with $a \leq 1/3 g^2T$ in this work, which
means that the CPU time taken to measure $\NCS$ is negligible for all
the most numerically intensive cases.

\begin{figure}[p]
\centerline{\epsfxsize=4in\epsfbox{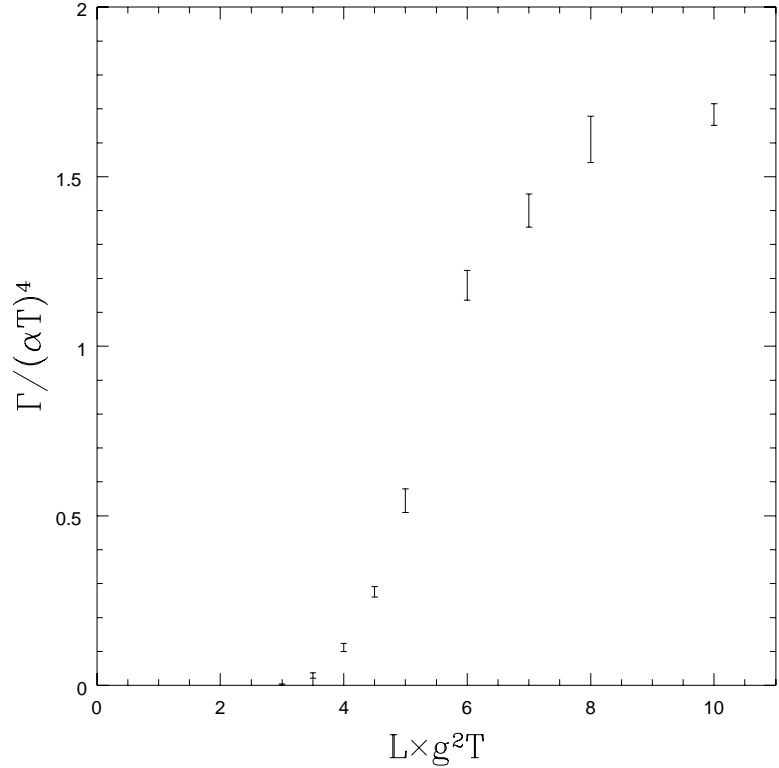}}
\caption{\label{Vol_dependence} Volume dependence of the sphaleron rate
in cubic volumes $L$ on a side with periodic boundary conditions and $a
= 1/2g^2 T$ ($\beta=8$).  Large
volume behavior is obtained by $L=8/g^2T$.  Around $L=5/g^2T$ the
sphaleron rate falls off abruptly and it is virtually zero already at
$L=3 / g^2T$.}
\end{figure}

\begin{table}[p]
\centerline{\begin{tabular}{|c|c|c|} \hline
$N$ & $L \times g^2 T$ & $\Gamma / \alpha^4 T^4$ \\ \hline
6   &  3.0   &  $0.0023 \pm 0.0016$ \\ \hline
7   &  3.5   &  $0.029 \pm 0.008$ \\ \hline
8   &  4.0   &  $0.112 \pm 0.012$ \\ \hline
9   &  4.5   &  $0.276 \pm 0.015$ \\ \hline
10  &  5.0   &  $0.545 \pm 0.035$ \\ \hline
12  &  6.0   &  $1.18  \pm 0.04$ \\ \hline
14  &  7.0   &  $1.40  \pm 0.05$ \\ \hline
16  &  8.0   &  $1.61  \pm 0.07$ \\ \hline
20  & 10.0   &  $1.68  \pm 0.03$ \\ \hline \end{tabular}}
\caption{\label{Vol_table} Volume dependence of the sphaleron rate for an
$a=1/2g^2T$ lattice.  The volume dependence is very strong at $L=4/g^2T$
but is consistent with zero above $L=8/g^2T$.}
\end{table}

\begin{figure}[p]
\centerline{\epsfxsize=4.3in\epsfbox{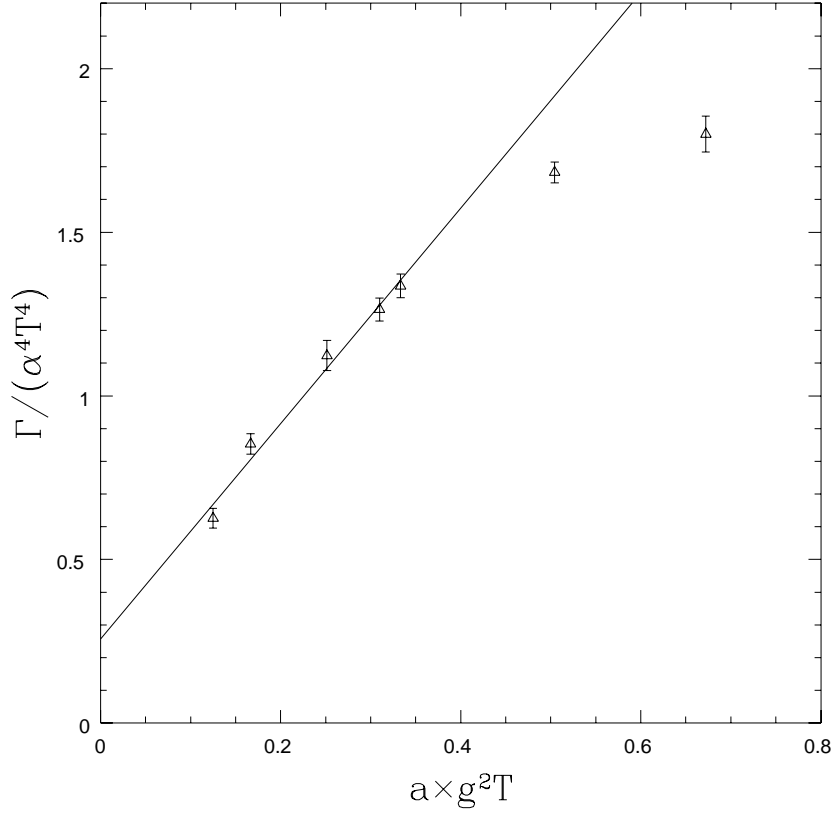}}
\caption{\label{results} Chern-Simons number diffusion constant plotted
against lattice spacing $a$, and a linear fit for the last 5 points.  
Around $a \sim 0.5/g^2T$ the dependence is
somewhat weak, but at larger $a$ a rapid falloff
is evident.}
\end{figure}

\begin{table}[p]
\centerline{\mbox{\begin{tabular}{|c|c|c|c|c|c|}\hline
$a\times g^2T$  & $\; \; \beta \; \;$ & $\; \; N \; \;$ 
	& time/$a$ & $\Gamma/(\alpha T)^4$ & 
	same, converting using $\betaL$\\ \hline
1.00    &  4  &  10  & 40000    & $2.123 \pm 0.075$ & $3.78 \pm 0.12$
\\ \hline
0.67    &  6  &  16  & 58000    & $1.800 \pm 0.055$ & $2.66 \pm 0.08$
\\ \hline
0.50    &  8  &  20  & 157000   & $1.683 \pm 0.032$ & $2.25 \pm 0.04$
\\ \hline
0.33    & 12  &  32  & 170000    & $1.336 \pm 0.036$ & $1.62 \pm 0.04$
\\ \hline
0.31    & 13  &  32  & 188000   & $1.264 \pm 0.035$ & $1.51 \pm 0.04$
\\ \hline
0.25    & 16  &  40  & 105000   & $1.123 \pm 0.046$ & $1.30 \pm 0.05$
\\ \hline
0.167   & 24  &  60  & 234400   & $0.853 \pm 0.031$ & $0.94 \pm 0.03$
\\ \hline
0.125   & 32  &  80  & 202000   & $0.626 \pm 0.030$ & $0.67 \pm 0.03$
\\ \hline
\end{tabular}}}
\caption{Sphaleron rate $\Gamma$ as a function of lattice spacing,
$\beta = 4 / g^2 aT$.  On the finer lattices an approximately linear
dependence on $a$ (or $1/\beta$) becomes apparent.  For those who prefer
it we also include $\Gamma$ converting to physical units using the naive
relation between $\betaL$ and $a$.  \label{table1}}
\end{table}

We should also be sure to use a large enough volume to eliminate finite
volume systematics (effectively, to perform the $V \rightarrow \infty$
limit in Eq. \ref{def_of_Gamma}).  To do so we measure $\Gamma$ as a
function of $L=V^{1/3}$ at a fixed lattice spacing, $a=1/2g^2T$.  The
result is shown in Figure \ref{Vol_dependence}.  The dependence on $L$
is in general accordance with the results of \cite{AmbKras}, except at
small volumes where ours go to zero and theirs do not because of the
spurious UV contributions in their definition of $\NCS$.  We will use
$L=10 / g^2 T$ ($N=2.5 \beta$) for all large volume results.

\pagebreak

Our results at large volumes are presented in Table \ref{table1} and
plotted in Figure \ref{results}.  These constitute the main result of
this paper.  While $\Gamma$ is not a strong function of lattice spacing
around $a \simeq 0.5/g^2 T$ ($\beta \simeq 8$), it then turns over and
falls roughly linearly in $a$ at finer lattices.  This indicates that
the lattice spacing needed before the ASY scaling behavior sets in is
around $\beta=12$ ($a=1/3 g^2T$).

If we insist on believing that the small $a$ scaling behavior is of form
$c_1 + c_2 g^2 aT$ with $c_2$ representing a correction to scaling, then
a fit to the points with $a \leq 1/3 g^2 T$ ($\beta \geq 12$) gives 
$c_1 = 0.257 \pm .044$ and $c_2 = 3.29 \pm .18$ ($\xi^2/\nu = 5.4/3$).
The ``correction'' to scaling only becomes subdominant below 
$a = 1/12.8 g^2 T$ ($\beta > 51$).  The original motivation for
believing in a finite small $a$ limit for $\Gamma$ is the belief that
the UV lattice behavior is not important to the infrared dynamics.  Such
an enormous correction to scaling contradicts this belief.  This makes
it very difficult to reconcile our data with a finite small $a$ limit
for $\Gamma$.  

\begin{figure}[t]
\centerline{\epsfxsize=4.4in\epsfbox{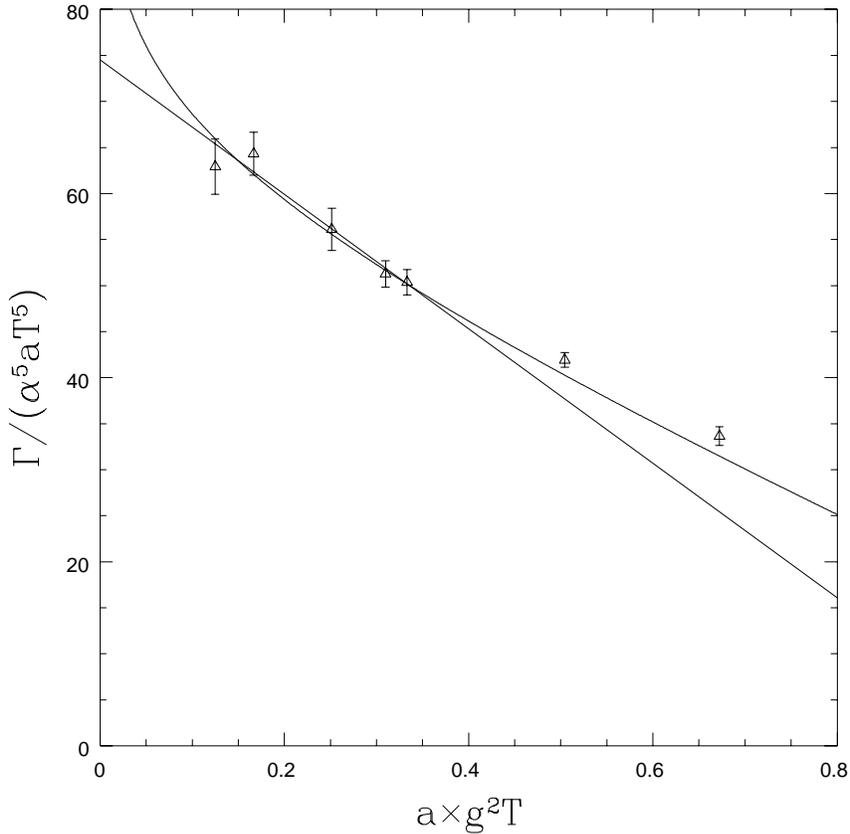}}
\caption{\label{otherway} $\Gamma/a$ against $a$.  The straight line fit
assumes the ASY scaling argument, the curved fit includes B\"{o}deker's
logarithmic correction to this scaling argument; each are based on the 5
highest points.}
\end{figure}

On the other hand, if ASY are correct, it makes more sense to plot our
results as $\Gamma / (\alpha^5 a T^5)$, as we do in Figure
\ref{otherway}.  The extrapolation to small $a$ looks much better
behaved here.  If we fit $\Gamma$ to the form $\Gamma = \alpha^5 aT^5 
(c_1 + c_2 g^2 aT)$ we get $c_1=74.5 \pm 3.4$ and $c_2 = -73 \pm 12$
($\xi^2 / \nu = 1.6/3$), which means the correction to scaling comes of
order 1 at $a=g^2 T$ ($\beta = 4$).

In fact, as B\"{o}deker has shown, we should not expect a
finite intercept in this figure, there should be a weak logarithmic
divergence as $a \rightarrow 0$ \cite{Bodeker}.  
In the continuum its amplitude is \cite{Bodek_paper}
\beq
{\rm log \; part \; of \;}\Gamma = (10.7 \pm .7)
	\frac{g^2 T^2}{m_D^2} \log \left( \frac{m_D}{g^2T} \right)
	\alpha^5 T^4 \, .
\eeq
The appearance of $1/m_D^2$ is changed somewhat on the lattice; $m_D^2$
is reduced by a weighted average over $k$ of the group velocity under
lattice dispersion relations, which is about $0.68$ \cite{Arnoldlatt}.
Combining this with the expression for $m_D^2$ on the lattice,
Eq. (\ref{mD_is}), gives 
\beq
{\rm log \; part \; of \;}\Gamma = \left[ 7.8 \pm .5 \right]
	 \log \left( \frac{1}{g^2 a T} 
	\right) \alpha^5 a T^5 \, .
\eeq
Strictly speaking such logarithmic behavior only pertains in the regime
where $\log(1 / g^2 aT) \gg 1$, which is probably not satisfied at any
conceivable lattice spacing.  Nevertheless we perform the fit to see
what happens.
The fit in Fig. \ref{otherway} which 
curves assumes $\Gamma = {\rm log \; part \; 
of \;} \Gamma + \alpha^5 a T^5 (c_1 + c_2 g^2 aT)$, which still has only
two free fitting parameters.  The best fit value is $c_1 = 54.6 \pm 3.4$,
$c_2=-39 \pm 12$ ($\xi^2/\nu = 2.1/3$).  In this case the leading $c_1$
behavior dominates over the scaling correction out to $a= 1.4 g^2 T$
($\beta = 3$).  Note that the data do not yet justify either believing
or disbelieving in the log behavior.  This is because the coefficient of
the log is numerically small.

\section{Small volumes}

\begin{figure}[t]
\centerline{\epsfxsize=3.0in\epsfbox{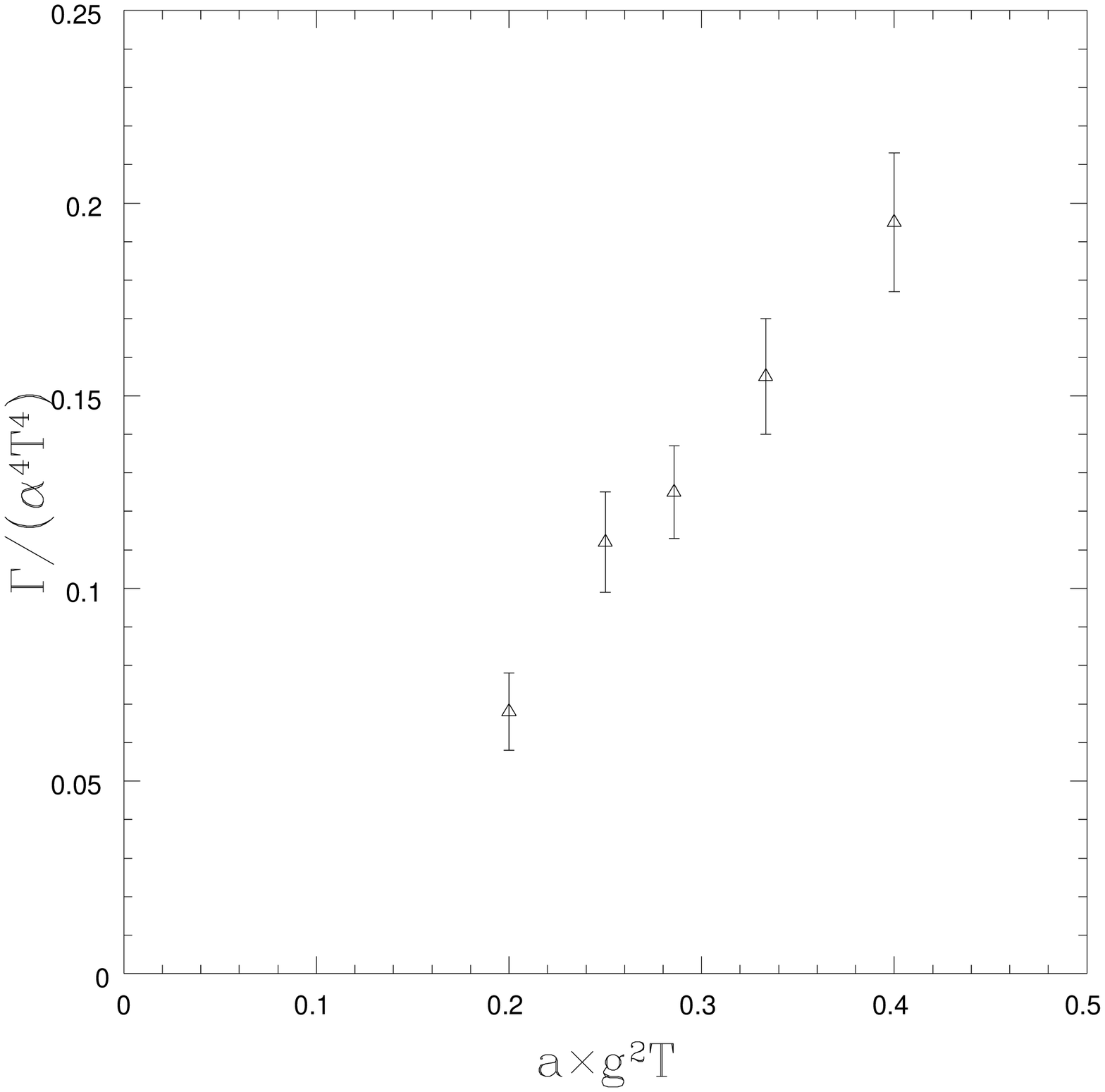} \hspace{0.3in}
\epsfxsize=3.0in\epsfbox{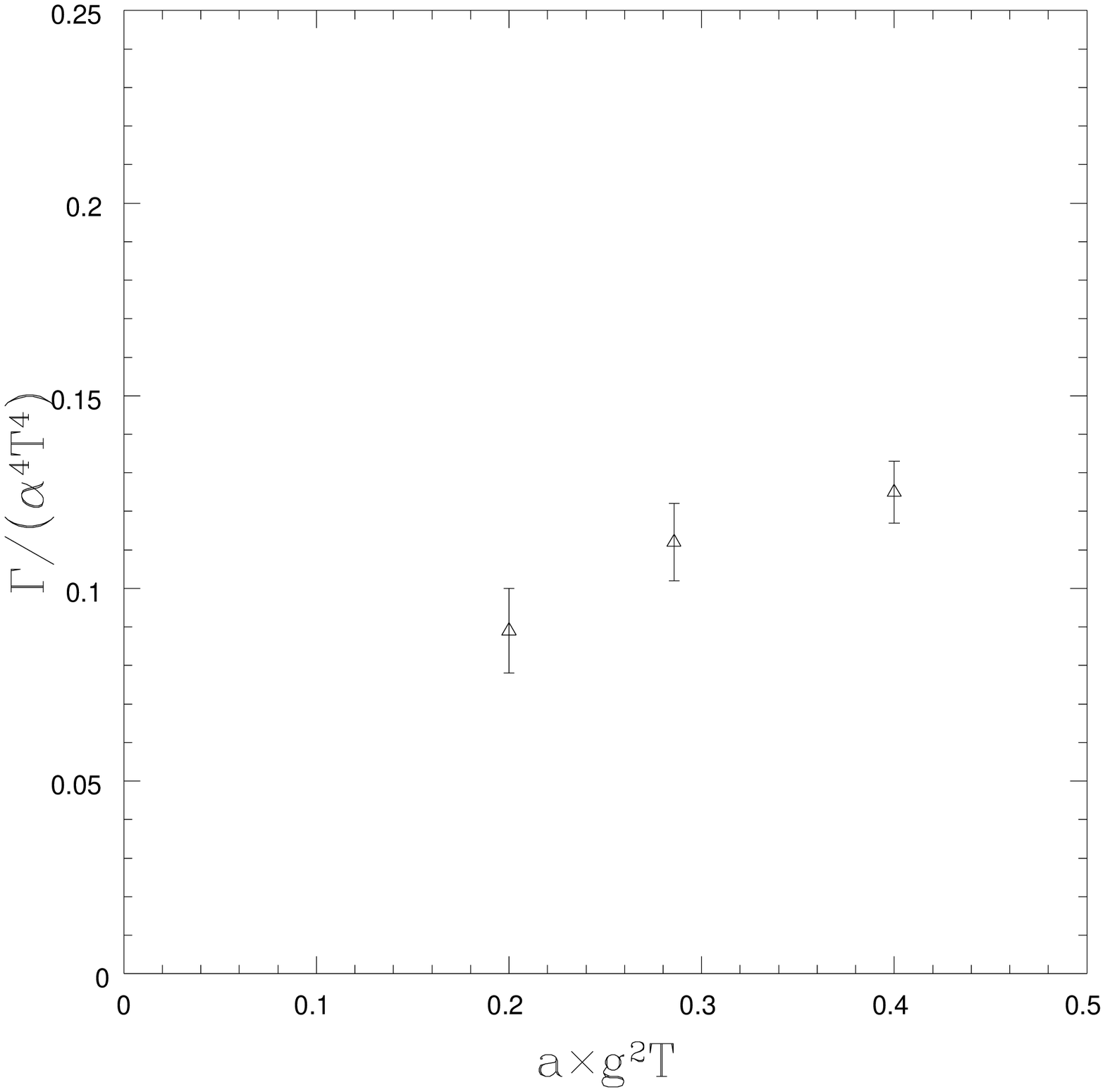}}
\caption{\label{small_vol} Dependence of $\Gamma$ on $a$ in a small
volume, $L \simeq 4 / g^2 T$.  
Left, the results Ambj{\o}rn and Krasnitz reported in
\protect{\cite{AmbKras2}}.  Right, results when more care is taken to
keep the lattice volume fixed in physical units.  The overly strong $a$
dependence in the Ambj{\o}rn-Krasnitz data is due to small changes in
the physical volume, which $\Gamma$ depends on very strongly.}
\end{figure}

Ambj{\o}rn and Krasnitz report another puzzling problem with the ASY
picture \cite{AmbKras2}.  
They analyzed the dependence of $\Gamma$ on lattice spacing at
a fixed, small volume, small enough that $\Gamma$ is much smaller than
its large volume limit.  The idea was that, in such a small volume, 
$\NCS$ fluctuates about an integer value with occasional, abrupt changes
from integer to integer, which can be identified even with the naive
definition of $\NCS$.  Therefore, $\NCS$ can be tracked topologically
without needing a true topological definition.
Their results, replotted here in Figure \ref{small_vol},
are puzzling.  Not only is there a strong lattice spacing dependence in
evidence; it is too strong.  The rate appears to fall off {\em faster}
than $\Gamma \propto a$, and certainly faster than the rate falls off in
large volumes, where the corrections to scaling at finite $a$ make the
$a$ dependence somewhat weaker than linear.

To see why this result does not jive with the ASY picture, we will
review again the basic ASY argument.  Examining the propagator for the
gauge field at momentum $p$ gives an equation of motion for $A(p)$
which, viewed on suitably long time scales,
is approximately
\beq
\frac{d^2 A(p)}{dt^2} + \frac{\pi m_D^2}{4 p} \frac{dA(p)}{dt} 
	= - p^2 A(p) \, . 
\eeq
The damping term, with the single derivative, is more and more important
as $p$ gets smaller.  We expect that, in a constrained volume, the field
configurations responsible for changing $\NCS$ are spatially smaller
than in a large volume.  Therefore they are composed of excitations of
larger $p$, and should be more weakly damped.  In particular it should
take a larger value of $m_D^2$ (smaller $a$) 
for the overdamped regime to apply, and
the corrections to linear in $a$ scaling should be larger.

The problem lies in the data.  Ambj{\o}rn and Krasnitz worked in a
lattice with $N = \betaL$, which at the unimproved level means 
$L = 4 / g^2 aT$.  However, they used the
tree relation between the lattice spacing and the reciprocal temperature
$\betaL$.  As can be seen from Eq. \ref{latt_improvement}, the true
lattice spacing is larger than the tree relation indicates, and the
error is worse as the lattice is made coarser.  Therefore, their larger
$a$ lattices possessed more physical volume than their smaller $a$
lattices.  The effect is fairly small; the difference in linear
dimension between their $\betaL=10$ and 
$\betaL=20$ lattices is only about
$3\%$.  However, as Fig. \ref{Vol_dependence} shows, $\Gamma$ is a very
strong function of volume in the regime where they were working.
Around $N=\beta$ ($L = 4 / g^2 T$), the data in the figure give
roughly $d(\log \Gamma)/d(\log L) \simeq 10$, so a $3\%$ change in
length could lead to a $30\%$ change in $\Gamma$, which is significant.
In fact this effect could be as large as or larger than the $a$
dependence due to hard thermal loop dynamics.

To fix this problem we recompute $\Gamma$ in a fixed, small volume, but
using the improved relation between the lattice spacing and the
reciprocal temperature.  We choose to use a volume equivalent (at the
improved level) to the volume of Ambj{\o}rn and Krasnitz' finest volume,
$\betaL=20$ and $N=20$, which, using the improved relation,
gives $L=4.114 / g^2 T$.  The results are presented in
Figure \ref{small_vol}, and include a re-calculation of the finest
lattice point, using a topological definition of $\NCS$ (rather than
counting integer winding changes by eye from the old ``unimproved''
$\NCS$ definition).  The new data show a lattice spacing dependence
similar to the large volume case, although to make a good comparison we
would need data at smaller $a$ (larger $\beta$).  The behavior of
$\Gamma$ in small volumes is in accord with what we expect if the ASY
argument is correct.

\section{Conclusion}

Chern-Simons number diffusion in pure Yang-Mills theory follows the
Arnold-Son-Yaffe scaling behavior, $\Gamma \propto a$; however it takes
a fairly fine lattice to demonstrate this in a convincing way.  The same
behavior, with similar corrections to scaling, is observed in small
volumes, but only after taking care to keep the physical volume constant
beyond leading (tree) level while the lattice spacing is varied.

The original goal of measuring $\Gamma$ on the lattice was to determine
the rate at which a baryon number excess, present before the electroweak
phase transition, would be erased.  We can do this with the results of
this paper by using Arnold's study of the matching between the lattice
and continuum theories \cite{Arnoldlatt}.  
He shows that $\Gamma$ on the lattice matches the continuum value when
\beq
(.68 \pm .20 )m_D^2({\rm latt}) = m_D^2({\rm continuum}) \, ,
\eeq
where the error is all systematic, but the error estimate is considered
generous \cite{Arnoldlatt}.  In the minimal standard model (MSM), at
leading order $m_D^2 = (11/6) g^2 T^2$, which for $g^2 = 0.42$ and using
Eq. (\ref{mD_is}), means the MSM value is obtained on a lattice with
$a = .157 / g^2 T$, or $\beta = 25.4$.  Our data at $\beta
= 24$ give $\Gamma = 0.85 \alpha^4 T^4$, with an error insignificant
compared to the $30\%$ error estimate in the lattice to continuum
matching procedure.  since $\Gamma \propto a$ is well satisfied in this
regime, our estimate for the Standard Model value of $\Gamma$ is
\beq
\Gamma({\rm MSM}) = (0.82 \pm .24) \alpha^4 T^4 \qquad {\rm or} 
	\qquad \Gamma = ( 45 \pm 13 ) \left( \frac{g^2 T^2}{m_D^2} \right)
	\alpha^5 T^4 \, ,
\eeq
where the latter form shows the correct parametric dependence on the
Debye mass.
This result is in good agreement with results obtained when classical
Yang-Mills theory is supplemented with particle degrees of freedom
\cite{particles} to induce the hard thermal loop effects.  Therefore the
diffusion constant for $\NCS$ in classical Yang-Mills theory is
consistent with both analytic expectations and numerical results
obtained by explicit inclusion of hard thermal loops.

\end{document}